\documentclass{aa}
\usepackage{graphicx}
\usepackage{natbib}
\usepackage{txfonts}
\begin{document}
   \title{Microlensing in the double quasar SBS1520+530}

   \author{E. R. Gaynullina\inst{1}
	\and R. W. Schmidt\inst{2}
	\and T. Akhunov\inst{1,3}
	\and O. Burkhonov\inst{3}
	\and S. Gottl{\"o}ber\inst{4}
	\and K. Mirtadjieva\inst{1,3}
	\and S. N. Nuritdinov\inst{1,3}
	\and I. Tadjibaev\inst{1,3}
	\and J. Wambsganss\inst{2}
	\and L. Wisotzki\inst{4}
   }

   \offprints{R. W. Schmidt, E-mail: rschmidt@ari.uni-heidelberg.de}

   \date{Draft \today}% Received xxx 2004; accepted yyy 2004}

   %\authorrunning{Gaynullina et al.}
   %\titlerunning{Maydanak monitoring of SBS1520+530}

   \institute{National University of Uzbekistan,
              Physics Faculty,
              Tashkent,
	      700174,
	      Uzbekistan
              \and
	      Astronomisches Rechen-Institut,
	      Zentrum f{\"u}r Astronomie der Universit{\"a}t Heidelberg,
	      M\"onchhofstra{\ss}e 12-14,
	      69120 Heidelberg,
	      Germany
	      \and
	      Ulugh Beg Astronomical Institute of the Uzbek Academy of
	      Sciences and Isaac Newton Institute of Chile, Uzbek
	      Branch, Astronomicheskaya 33, Tashkent, 700052, Uzbekistan
              \and
	      Astrophysikalisches Institut Potsdam,
	      An der Sternwarte 16,
	      14482 Potsdam,
	      Germany
	}

   \abstract{We present the results of a monitoring campaign of the
double quasar SBS1520+530 at Maidanak observatory from April 2003 to
August 2004. We obtained light curves in $V$ and $R$ filters that show
small-amplitude $\Delta m\approx 0.1$ mag intrinsic variations of the
quasar on time scales of about 100 days. The data set is consistent
with the previously determined time delay of $\Delta t=(130\pm3)$ days
by \citet{Burud02}. We find that the time delay corrected magnitude
difference between the quasar images is now larger by $(0.14\pm0.03)$
mag than during the observations by \citet{Burud02}. This confirms the
presence of gravitational microlensing variations in this system.

   \keywords{ gravitational lensing -- 
        dark matter  --
        quasars: individual: SBS1520+530 --
        cosmology: observations 
               }

   }

   \maketitle
%________________________________________________________________

\section{Introduction}

The broad absorption line (BAL) quasar SBS1520+530 ($z_{\rm q}=1.855$) was
discovered by \citet{Chavushyan97} as a gravitationally lensed double
quasar with an angular separation of 1\farcs56. The lensing galaxy was
detected by \citet*{Crampton98} using infrared adaptive optics imaging
at the Canada-France-Hawaii Telescope. \citet*{Faure02} observed the
lensing galaxy with the Hubble Space Telescope. \citet{Burud02} (in
the following B02) finally succeeded in obtaining both the redshift
$z_{\rm gal}=0.717$ (consistent with absorption lines first found by
\citealt{Chavushyan97}) of the lensing galaxy with a Keck obervatory
spectrum and the time delay of $\Delta t=(130\pm3)$ days between the
two quasar images using monitoring data from the Nordic Optical
Telescope (NOT). For this, B02 obtained an almost gapless lightcurve
of the object of about 800 days between February 1999 and May 2001. An
almost continuous coverage of the light curve of SBS1520+530 is made
possible by its high declination. Further photometry obtained at
Maidanak observatory on the system was published by
\citet*{Zheleznyak03}.

Interestingly, B02 found that by simply shifting the light curve of
image B backward by 130 days and correcting for the magnitude
difference $\Delta m=0.69$ mag of the images made the quasar light
curves align only approximately. They obtained a better match by
allowing for an additional linear trend, and by correcting for faster
variations using an iterative scheme \citep{Burud01}. B02 interpret
these additional variations to be probably due to microlensing
variability.

SBS1520+530 thus is one of the prime targets for microlensing studies
since it provides at once the prospects for long, uninterrupted light
curves with a known, relatively short time delay and known
microlensing variations. For this reason we decided to continue the
optical monitoring of this system.

\section{Observations}
\label{observations}

\begin{figure}
\begin{center}
\resizebox{\columnwidth}{!}{\includegraphics{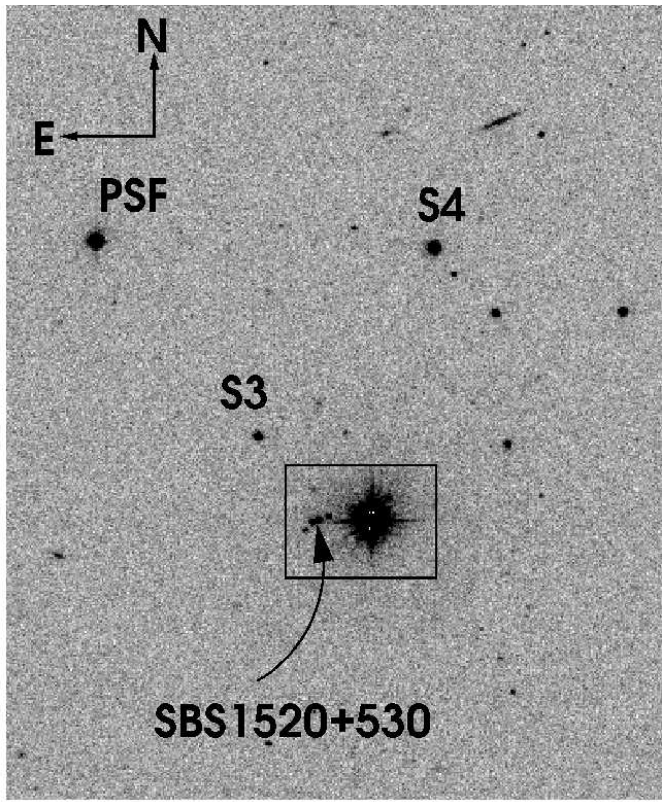}}
\end{center}
\caption{$R$-band image of SBS1520+530 obtained on May 7, 2003. The quasar
  images, the reference stars S3 and S4, and the star we use for the
  PSF are labelled. The field size is 2.8 arcmin$\times$3.5
  arcmin. The area marked with the box is shown in Fig.~\ref{zoom}.
}
\label{fullimage}
\end{figure}

We report here on our observations of SBS1520+530
(Fig.~\ref{fullimage}) with the 1.5\,m AZT-22 telescope in Maidanak,
Uzbekistan \citep{Ehgamberdiev00}, between April 25 and October 12 in
2003, and again between January 16 and August 22 in 2004. In these
intervals we observed the object almost daily whenever the weather
permitted it. The $V$ and $R$-filters provide photometry in the
Johnson-Cousins system. We observed on average four frames per night
with an exposure of 3.5 min in $V$ and 3 min in $R$ with a median
seeing of 1.07 arcsec.

In total we obtained 116 nights in the $V$ band and 191 nights in the $R$
band. After selecting for good seeing (better than 1.4 arcsec
full-width at half maximum) and low sky-background we finally used 80
nights in the $V$-band and 123 nights in the $R$-band. The observations
were made with the BROCAM CCD detector with a pixel scale of
0.26 arcsec. Bias correction and flat fielding of the images were done
using standard IRAF software.

\section{Photometry}
\label{photometry}

\begin{figure}
\begin{center}
\resizebox{\columnwidth}{!}{\includegraphics{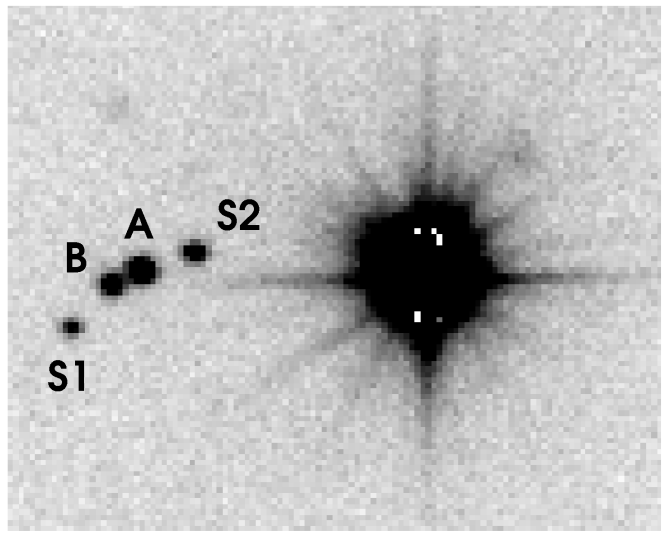}}
\end{center}
\caption{$R$-band zoom of the central part of SBS1520+530. The field
size is 30 arcsec$\times$25 arcsec. North is up and East is to the
left. The bright foreground star next to the two quasar images A and B
is saturated in the centre. S1 and S2 are also foreground stars. The
lens galaxy ($m_{R, {\rm gal}}\approx 21.6$ mag, \citealt{Crampton98})
is too faint to be seen in this image.}
\label{zoom}
\end{figure}

A special property of the SBS1520+530 system is its location within 14
arcseconds of a bright 12th magnitude star (Fig.~\ref{zoom}). This is
great for adaptive optics studies of the system because a bright
reference star is at hand \citep{Crampton98}. For photometry,
however, the bright halo and the diffraction pattern caused by the
star on the CCD needs to subtracted carefully.

In order to do this, we followed the method described by
\citet{Zheleznyak03};  
we extracted the western half of the star and subtracted it from the
eastern part where SBS1520 is situated. This procedure also
efficiently subtracts the light due to the horizontal diffraction spike
that extends towards the double quasar.

Photometry on the quasar components A and B was performed using the
DAOPHOT package \citep{Stetson87}. We chose this method because
it is well-suited to the mildly asymmetric and time-variable AZT-22
point spread function (PSF). Because of the variable PSF it was not
possible to use the image subtraction technique
\citep{Alard98,Alard00}. The quoted error bars are the standard $1\sigma$
errors determined by DAOPHOT. Since there are four point sources (the
quasar components A and B, and the stars S1 and S2) crowded in a
rather small region, we had to fit the positions and magnitudes of all
four components at the same time.
Absolute quasar magnitudes were calibrated using the
brightness of the reference star S3 (see Fig.~\ref{fullimage}) in the
$V$-band ($m_{V}=(17.37\pm0.02)$ mag) and $R$-band ($m_{
R}=(17.18\pm0.02)$ mag) determined by \citet{Zheleznyak03}. The star
we used as a template to model the point spread function in DAOPHOT is
marked with ``PSF'' in Fig.~\ref{fullimage}.

In our analysis we ignore the presence of the lensing galaxy near
image B because it is too faint to be detected in our images
($m_{R,{\rm gal}}=21.6$ mag, \citealt{Crampton98}). If all of the
galaxy light contributes to our magnitude estimate of quasar image B,
the measured brightness would increase by a constant offset of 0.08
mag (assuming $m_{R,{\rm B}}=18.8$ mag).

\begin{figure}
\begin{center}
\rotatebox{270}{\resizebox{!}{\columnwidth}{\includegraphics{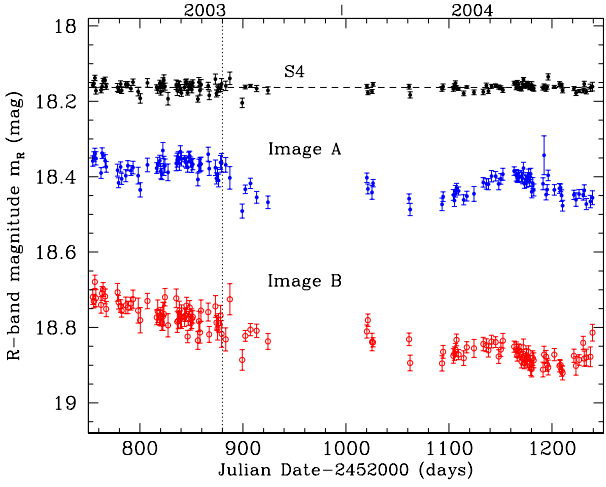}}}
\end{center}
\caption{$R$-band light curves of the two quasar images A and B in
  SBS1520+530 and the reference star S4. For clarity the magnitude for
  image B was shifted by -0.4 mag, S4 was shifted by +2.4 mag. The
  dotted vertical line indicates the day with the Julian
  Date-2452000=880 when the mirror was cleaned. The dashed horizontal
  line shows the magnitude of the reference star S4 $m_{\rm S4}=18.16$
  mag.}
\label{lightcurve}
\end{figure}

\begin{figure}
\begin{center}
\resizebox{\columnwidth}{!}{\includegraphics{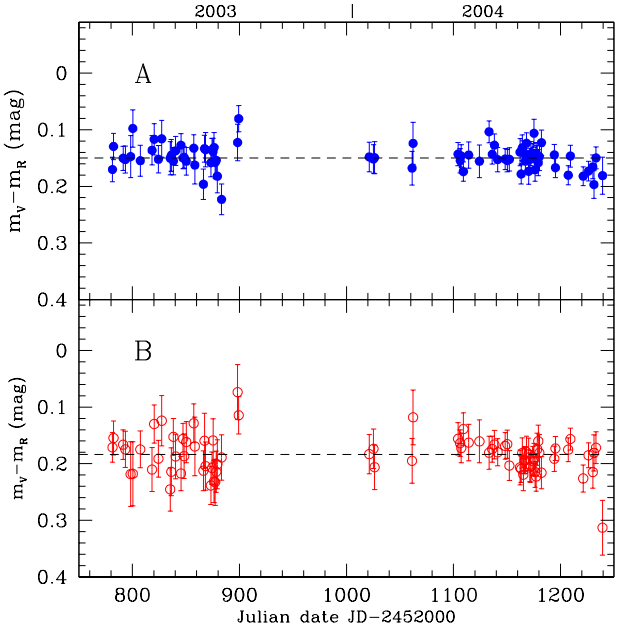}}
\end{center}
\caption{$V-R$ colour curve for the quasar images A (top) and B
  (bottom). The best-fitting value for each quasar image is plotted
  with a dashed line.}
\label{colour}
\end{figure}

\section{Results}
\label{apo_results}

\subsection{Light curves}

We present the results of the $R$-band photometry of SBS1520+530 in
Fig.~\ref{lightcurve} and Table~1\footnote{Table 1 is being made
available electronically at CDS. Columns 1-4 contain the A and B
magnitudes with errors, Column 5 contains the Julian Date of the
observations.}. The light curves of the two quasar components A and B
are plotted together with the light curve of an additional reference
star S4 (see Fig.~\ref{fullimage}, $m_{R}=18.16$ mag). This shows that
both S3 (which is used for the absolute magnitude calibration) and S4
do not vary.

Both quasar components show low-amplitude $\Delta m\approx 0.1$ mag
variations on time scales of about 100 days. As we will show in the
next section, the overall similarity of the two quasar image light
curves is due to a gradual brightness decrease of the lensed quasar.
On the day with Julian Date-2452000=880 the telescope mirror was
cleaned. This led to a large sensitivity improvement that visibly
improved the accuracy of the quasar brightness measurements,
especially for the fainter image B.

In Fig.~\ref{colour} and Table 2\footnote{Table 2 is being made
available electronically at CDS. Columns 1-4 contain the $V-R$ colours
for images A and B with errors, Column 5 contains the Julian Date of
the observations.}  we show the difference between our $V$-band and
$R$-band light curves of SBS1520+530. We find an average $V-R$ colour
$m_{V}-m_{R}=0.15$ mag for image A and $m_{V}-m_{ R}=0.18$ mag for
image B. We do not find any evidence for significant colour variations
during our observing interval. The small difference $\Delta (V-R)
\approx 0.03$ mag of the $V-R$ colour between the quasar images
indicates the presence of a small level of differential reddening
along the light paths.

\subsection{Time delay}
\label{timedelay}

A wide variety of algorithms have been developed for the determination
of time delays in gravitational lens systems \citep[e.g.,
][]{Kundic97,Burud01,GilMerino02}. We choose here a strightfoward
linear interpolation scheme because our light curve mainly consists of
two frequently sampled observing intervals. The unfrequently sampled
gap in the middle cannot be confidently interpolated with any
method. In detail we use the following recipe:
\begin{enumerate}
\item The B light curve is shifted by the time delay $\Delta t$ to be tested.
\item One of the two light curves is linearly interpolated to match
  the observing dates of the other light curve.
\item Only gaps that are less than 40 days (including the gaps between
  Julian Date-2452000=1000 and 1100) are interpolated, no difference
  is calculated for larger gaps because they can introduce a false
  signal due to sparse sampling of the light curve.
\item For the remaining $N$ days of overlap, the weighted difference
  $\Delta m=<m_{\rm A}- m_{\rm B}>$ is determined and the goodness of fit
estimator
\begin{equation}
\chi^2_{\nu}=\frac{1}{N-2}\sum_{\rm i=1}^{N} \frac{\left(m_{\rm A}
  (t_i)-m_{\rm B} (t_i+\Delta t)-\Delta m\right)^2}
{\sigma^2_{\rm A}+\sigma^2_{\rm B}}
\end{equation}
(corresponding to the time delay $\Delta t$) is calculated.
\end{enumerate} 
For the linear interpolation the errors are added in quadrature. The
number of degrees of freedom is $\nu=N-2$ because there are two free
paramaters: time delay and magnitude shift. The $1/(N-2)$ factor
penalizes solutions with a small number of overlap days.

The best-fitting time delay can be determined by calculating
$\chi^2_{\nu}$ values for a range of time delays, and by choosing the
time delay with the lowest $\chi^2_{\nu}$ value. In order to determine
the associated measurement uncertainties we used 10000 bootstrap
resamplings of the observed light curve, smoothed by a triangular
filter with a full width of 20 days \citep[e.g., ][]{Kundic97}. For
each smoothed resampling the best-fitting time delay was determined
for time delays between 0 and 220 days (image A leading). This
procedure is robust with respect to the size of the filter.  We always
interpolated image A because in this case the sparse sampling of our
light curve between Julian Date-2452000=900 and 1100 has a smaller
impact on the determined limits. We have verified that interpolating
image B does not change the answers. It would only weaken the
determined constraints.

\begin{figure}
\begin{center}
\resizebox{\columnwidth}{!}{\includegraphics{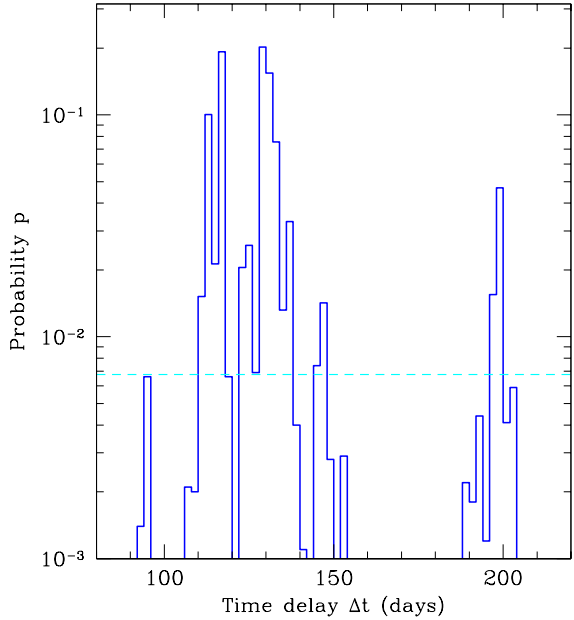}}
\end{center}
\caption{ Probability $p$ for time delays based on our SBS1520+530
light curve. The probability was calculated from $10000$ Monte-Carlo
realizations of the light curve, and by interpolating quasar A each
time. 95 per cent of the probability is contained in the four separate
regions above the dashed line.}
\label{deltat}
\end{figure}

The result of the Monte-Carlo resampling is shown in
Fig.~\ref{deltat}. For two-day bins of the time delay $\Delta t$ the
number of Monte-Carlo light curves with a best-fitting time delay in
the given range has been calculated. The probability $p$ of each bin
was calculated by dividing by the total number of resamplings. We find
that there are several time delays that are consistent with our data;
the 95 per cent confidence region consists of four separate
regions. For these sub-regions we can calculate the average time delay
and the standard deviation, yielding $\Delta t=(115.4\pm2.1)$ days,
$(130.5\pm2.9)$ days, $(146.3\pm1.0)$ days and $(198.8\pm1.1)$
days. The regions carry 33 per cent, 54 per cent, 2 per cent and 6 per
cent, respectively, of the statistical weight.

The sub-region with the largest statistical weight is consistent with
the time delay $\Delta t=(130\pm3)$ days found by B02. For this
time delay our data require an offset of $\Delta m=-0.83$ mag. In the
remainder of this paper we will use this time delay for our
data.

\begin{figure}
\begin{center}
\resizebox{\columnwidth}{!}{\includegraphics{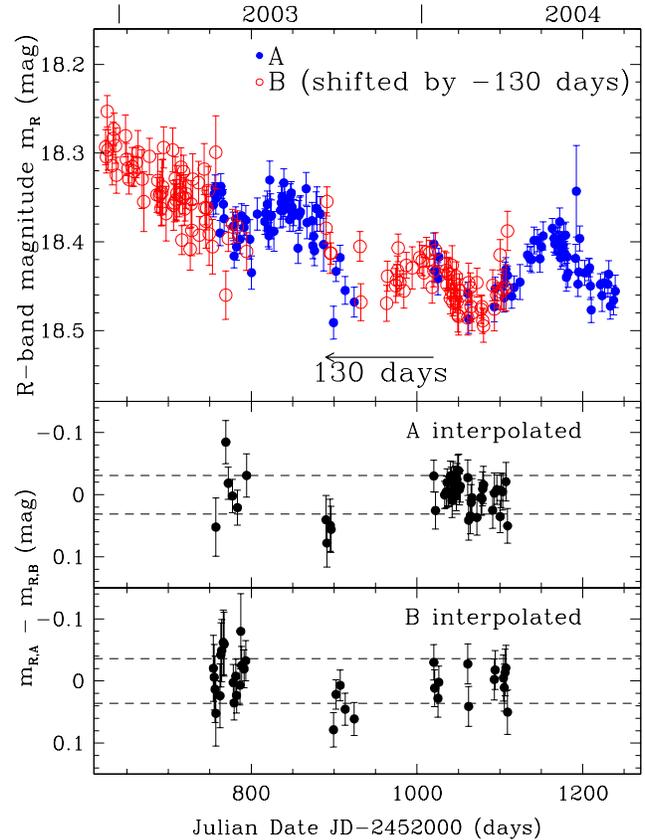}}
\end{center}
\caption{{\bf Top panel}: $R$-band light curves of the two quasar images
  A (filled circles) and B (open circles). The B light curve has been
  shifted by $-130$ days, as
  indicated by the arrow, and by $-0.83$ mag.
  {\bf Bottom panels}: Difference light curve between the shifted and
  interpolated quasar images. This was calculated in both ways: by
  interpolating image A or by interpolating image B (see text).
}
\label{delayed}
\end{figure}

\subsection{Microlensing}

In the top panel of Fig.~\ref{delayed} we show the quasar A and B
light curves in one plot, where image B was shifted to the left by the
time delay of $130$ days and up by the magnitude offset of $-0.83$ mag
(see Sect.~\ref{timedelay}). The composite light curve has no large
gaps. It shows that the quasar has been going through a series of
three small $\Delta m\approx 0.1$ mag brightness variations that each
lasted about 100 days.

Microlensing in the lens galaxy would only affect one of the light
paths to the quasar and could thus be detected as a residual light
curve difference \citep[e.g., ][]{Schmidt98,Wambsganss00}. In order to
study whether microlensing variations are present in our data, we
calculated the difference between the two observed light curves for
the time delay $\Delta t=130$ days found by B02 and the magnitude
offset of $\Delta m=-0.83$ mag.

To calculate the difference light curve, quasar B was shifted in time
and magnitude. The rest of the procedure is identical to the one
described in Sect.~\ref{timedelay}; we calculated the difference by
linearly interpolating the light curves of quasar A or B. The light
curves were interpolated whenever the gap was less than 40 days. No
difference was calculated for larger gaps. The error bars were added
in quadrature. The resulting two difference light curves are plotted
in the bottom two panels of Fig.~\ref{delayed}.

It can be taken from these plots that we do not detect a significant
difference between the two quasar light curves. We can calculate the
goodness-of-fit estimator for the null hypothesis that there is no
difference between the light curves on the $N$ days for which the
difference was calculated (yielding $\nu=N-1$ degrees of freedom):
\begin{equation}
\chi_{\nu}^2=\frac{1}{N-1}\sum_{\rm i} \frac{\Delta m_{\rm
    i}^2}{\sigma_{\rm i}^2}.
\end{equation}
This calculation yields $\chi_{\nu}^2=1.0$ regardless of whether image
A or B is interpolated, indicating that there is excellent agreement
between the light curves. The data are formally compatible at a
probability of 47 per cent with the null hypothesis. This
$\chi_{\nu}^2$-procedure ignores a possible temporal correlation of
the data, but there is also no evidence in Fig.~\ref{delayed} for such
a correlation.

Assuming that the data have a Gaussian scatter around zero we can
determine the standard deviation
\begin{equation}
\sigma^2=\frac{1}{N-1}\sum_{\rm i} {\Delta m_{\rm i}^2}.
\end{equation}
This yields $\sigma=0.03$ mag if image A is interpolated and
$\sigma=0.04$ mag if image B is interpolated. These values are
indicated in Fig.~\ref{delayed} (dashed lines). They are of the same
magnitude as the error bars, again showing that the difference light
curve is consistent with being entirely due to measurement
uncertainties.

\begin{figure}
\begin{center}
\rotatebox{270}{\resizebox{!}{\columnwidth}{\includegraphics{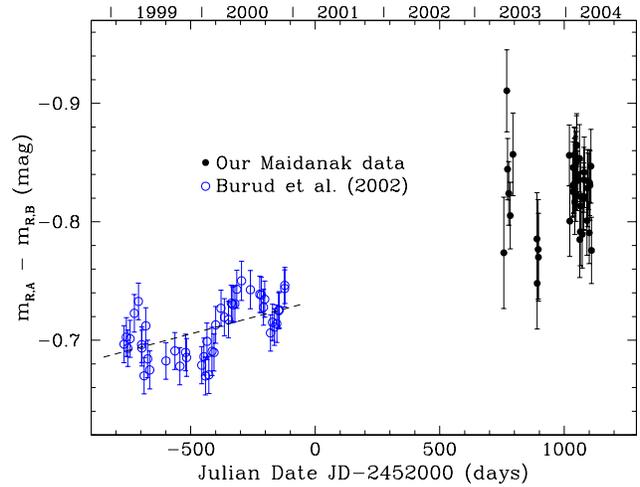}}}
\end{center}
\caption{ Composite of the difference light curves based on our
Maydanak data (filled circles, see Fig.~\ref{delayed}) and the data
published by \citet{Burud02} (open circles, the linear trend
determined by them is shown with a dashed line). In both cases image A
was linearly interpolated.}
\label{ml-plot}
\end{figure}

In their earlier data taken between February 1999 and May 2001, B02
did find a difference between the light curves of the two quasar
images in this system. Applying our procedure to the light curves in
their table 2, we can also calculate the difference light curve of
their data. The result is shown with in Fig.~\ref{ml-plot} (open
circles) together with our difference light curve (filled circles)
from the first of the two bottom panels in Fig.~\ref{delayed} (image A
was linearly interpolated). The difference curve in this figure is
plotted without offset.

Fig.~\ref{ml-plot} shows that B02 observed a coherent and highly
significant difference light curve with a maximum amplitude of $\Delta
m\approx0.08$ mag. Since the time of the observations by B02 the
magnitude difference between the quasar images has increased by
$0.14\pm0.03$ mag (see the discussion in Sect.~\ref{discussion}). B02
already identified a linear trend (dashed line) in their data. Our
data are consistent with this linear trend having continued until the
epoch of our observations. 

Although the exposure times for the Maidanak and the B02 data are
similar, the error bars of the Maidanak data are larger than the error
bars obtained by B02. The main reason for this are the different
telescope apertures (1.5m at the AZT-22 vs 2.5m at the NOT). In
addition, however, the AZT-22 transmission was reduced by about one
magnitude before the mirror was cleaned on Julian Date-2452000=880
(see Fig.~\ref{lightcurve}).

We note that the long-term variation of the difference light
curve may even be slightly larger than shown in this plot because some
light from the lensing galaxy could be included in our magnitude
estimate of image B (see Sect.~\ref{photometry}). If all of the
lensing galaxy light were included, the Maidanak difference light
curve would have to be shifted upward by a constant offset of
-0.08\,mag to correct for the galaxy contribution.

\section{Summary and discussion}
\label{discussion}

We have presented $V$-band and $R$-band photometry of the gravitational
lens system SBS1520+530 taken at Maidanak Observatory in the years
2003 (April to October) and 2004 (January to August). During the
$\approx 500$ day observation period with 80 data points in $V$ and
123 data points in $R$ we find small amplitude intrinsic variations
($\Delta m\approx 0.1$ mag) on time scales of about 100 days in both
quasar images. The $V-R$ colour of the quasar is consistent with being
constant during the observed period.

Using linear interpolation of the quasar light curves, we have
determined the 95 per cent confidence region of time delays for our
data set. Due to gaps in the light curve, our data allow four separate
values of the time delay, the best one of which agrees with the time
delay of $(130\pm3)$ days found by B02. Image A is leading, which is
also consistent with lens models of the system
(\citealt{Asano00,Faure02};B02;\citealt{Zheleznyak03}).

Using the B02 time delay, we calculated the difference light curve
between the two quasar images. This shows that within the statistical
uncertainty the two quasar light curves are identical during our
observing interval (Fig.~\ref{delayed}). In the observations taken by
B02 between 1999 and 2001, a highly significant and variable
difference with an amplitude of $\Delta m\approx 0.08$ mag was present
for a fraction of the observing interval. Since then, the overall
$R$-band magnitude difference between the A and B light curves has
changed by $0.14\pm0.03$ mag, the difference being larger in our data
(Fig.~\ref{ml-plot}).

Any variable difference between the light curves of SBS1520+530 can be
interpreted as gravitational microlensing because other changes of the
source would be visible in both quasar images, delayed by the time
delay. In addition to the microlensing variability on short
time-scales ($\approx$ 100 days and less) found by B02, our data show
that there are also variations on longer time-scales of $\Delta
t\approx(100-1000)$ days. This overall level of microlensing
variations in SBS1520+530 appears comparable to variations seen in
other lens systems \citep[e.g., ][]{Hjorth02,Wyithe02,Schechter03}.

An exciting prediction for the microlensing effect of quasars is the
colour-dependence of the microlensing light curve in the vicinity of
caustics \citep{Wambsganss91}. In such a situation the difference
between the $V$ and the $R$ light curve could provide valuable clues
to the source structure of the quasar. We will continue to observe
SBS1520+530 from Maidanak observatory because frequent sampling of the
source remains crucial to derive limits on microlensing variability.
If colour variations associated with microlensing could be proven in
this system, there would be a strong case for parallel spectral
observations of the quasar (see also B02).

Since microlensing currently remains the only technique with the
promise to scan the continuum emission regions of quasars on
microarcsecond scales, SBS1520+530 should be viewed as a prime target
because of the combination of known quasar variability and the at
least occasional occurence of microlensing diagnostics at the same
time.

\begin{acknowledgements}

We thank B. P. Artamonov and V. N. Dudinov for useful advice on the
realization of our observations. We thank the former German Ambassador
to Uzbekistan, Dr. Martin Hecker, for his support of our
collaboration.
The Uzbek team thanks the AIP and the University of Potsdam for
hospitality during visits. SG and RWS thank the Ulugh Begh
Astronomical Institute for hospitality. This project was supported by
the German Research Foundation (DFG), grant 436 USB 113/5/0-1. We
also acknowledge support by the European Community's Sixth Framework
Marie Curie Research Training Network Programme, Contract No.
MRTN-CT-2004-505183 "ANGLES".

\end{acknowledgements}

\end{document}